\documentclass[twocolumn,english,aps,prl,groupedaddress,superscriptaddress]{revtex4}
\usepackage{amsmath}
\usepackage{amsbsy}
\usepackage{graphicx}
\usepackage{units}
\usepackage{xcolor} %colored text, \textcolor{color}{text}
\usepackage{soul} %stikeout lines, \st{text}
\usepackage{amsmath}
\usepackage{amsfonts}
\usepackage{mathrsfs}
\usepackage{amsbsy}
\usepackage{bm}
\usepackage{babel}

\begin{document}

\title{Stripe Antiferromagnetic Spin Fluctuations in SrCo$_{2}$As$_{2}$}

\author{W.~Jayasekara}
\affiliation{Ames Laboratory, Ames, Iowa, 50011, USA}
\affiliation{Department of Physics and Astronomy, Iowa State University, Ames, Iowa, 50011, USA}
\author{Y.~Lee}
\affiliation{Ames Laboratory, Ames, Iowa, 50011, USA}
\affiliation{Department of Physics and Astronomy, Iowa State University, Ames, Iowa, 50011, USA}
\author{Abhishek~Pandey}
\affiliation{Ames Laboratory, Ames, Iowa, 50011, USA}
\affiliation{Department of Physics and Astronomy, Iowa State University, Ames, Iowa, 50011, USA}
\author{G.~S.~Tucker}
\affiliation{Ames Laboratory, Ames, Iowa, 50011, USA}
\affiliation{Department of Physics and Astronomy, Iowa State University, Ames, Iowa, 50011, USA}
\author{A.~Sapkota}
\affiliation{Ames Laboratory, Ames, Iowa, 50011, USA}
\affiliation{Department of Physics and Astronomy, Iowa State University, Ames, Iowa, 50011, USA}
\author{J.~Lamsal}
\affiliation{Ames Laboratory, Ames, Iowa, 50011, USA}
\affiliation{Department of Physics and Astronomy, Iowa State University, Ames, Iowa, 50011, USA}
\author{S.~Calder}
\affiliation{Oak Ridge National Laboratory, Oak Ridge, Tennessee, 37831, USA}
\author{D.~L.~Abernathy}
\affiliation{Oak Ridge National Laboratory, Oak Ridge, Tennessee, 37831, USA}
\author{J.~L.~Niedziela}
\affiliation{Oak Ridge National Laboratory, Oak Ridge, Tennessee, 37831, USA}
\author{B.~N.~Harmon}
\affiliation{Ames Laboratory, Ames, Iowa, 50011, USA}
\affiliation{Department of Physics and Astronomy, Iowa State University, Ames, Iowa, 50011, USA}
\author{A.~Kreyssig}
\affiliation{Ames Laboratory, Ames, Iowa, 50011, USA}
\affiliation{Department of Physics and Astronomy, Iowa State University, Ames, Iowa, 50011, USA}
\author{D.~Vaknin}
\affiliation{Ames Laboratory, Ames, Iowa, 50011, USA}
\affiliation{Department of Physics and Astronomy, Iowa State University, Ames, Iowa, 50011, USA}
\author{D.~C.~Johnston}
\affiliation{Ames Laboratory, Ames, Iowa, 50011, USA}
\affiliation{Department of Physics and Astronomy, Iowa State University, Ames, Iowa, 50011, USA}
\author{A.~I.~Goldman}
\affiliation{Ames Laboratory, Ames, Iowa, 50011, USA}
\affiliation{Department of Physics and Astronomy, Iowa State University, Ames, Iowa, 50011, USA}
\author{R.~J.~McQueeney}
\affiliation{Ames Laboratory, Ames, Iowa, 50011, USA}
\affiliation{Department of Physics and Astronomy, Iowa State University, Ames, Iowa, 50011, USA}

\begin{abstract}
Inelastic neutron scattering measurements of paramagnetic SrCo$_{2}$As$_{2}$ at $T=5$ K reveal antiferromagnetic (AFM) spin fluctuations that are peaked at a wavevector of $\textbf{Q}_{\mathrm{AFM}}=(1/2,1/2,1)$ and possess a large energy scale.  These stripe spin fluctuations are similar to those found in $A$Fe$_{2}$As$_{2}$ compounds, where spin-density wave AFM is driven by Fermi surface nesting between electron and hole pockets separated by $\textbf{Q}_{\mathrm{AFM}}$.  SrCo$_{2}$As$_{2}$ has a more complex Fermi surface and band structure calculations indicate a potential instability towards either a ferromagnetic or stripe AFM ground state.  The results suggest that stripe AFM magnetism is a general feature of both iron and cobalt-based arsenides and the search for spin fluctuation-induced unconventional superconductivity should be expanded to include cobalt-based compounds.
\end{abstract}
\date{\today}
\maketitle

The $A$Fe$_{2}$As$_{2}$ compounds ($A=$ Ca, Sr, Ba) are itinerant antiferromagnets (AFMs) where spin-density wave ordering is driven by Fermi surface nesting between electron and hole pockets \cite{Johnston10}. The in-plane nesting vector $\textbf{Q}_{\mathrm{AFM}}=(1/2,1/2)$ describes a stripe AFM structure consisting of ferromagnetic (FM) chains of spins extending along the $[1,\bar{1}]$ direction with AFM alignment along $[1,1]$ [see Fig. 1(f)].  In Ba(Fe$_{1-x}$Co$_{x}$)$_{2}$As$_{2}$, electron-doping by the substitution of Co for Fe destabilizes the stripe AFM ordering by shrinking (enlarging) the hole (electron) pockets and detuning the nesting condition.  Ultimately, the suppression of stripe AFM ordering upon Co substitutions of a few percent allows a superconducting ground state to appear in the presence of substantial spin fluctuations at $\textbf{Q}_{\mathrm{AFM}}$. Further Co substitutions ($x > 12\%$) lead to a complete suppression of both stripe spin fluctuations \cite{Sato09,Sato11} and superconductivity \cite{Sefat08,Leithe08,Ni08}.

The $A$Co$_{2}$As$_{2}$ compounds with a full replacement of Fe by Co have garnered little attention.  Initial experiments on BaCo$_{2}$As$_{2}$ \cite{Sefat09} and SrCo$_{2}$As$_{2}$ \cite{Pandey13} describe these materials as metals with enhanced paramagnetic susceptibility and no magnetic ordering or superconductivity down to 2 K.  Band structure calculations find a large density-of-states at the Fermi energy that is proposed to drive a ferromagnetic instability and enhanced paramagnetism \cite{Sefat09,Pandey13}.  Recent angle-resolved photoemission spectroscopy (ARPES) data on BaCo$_{2}$As$_{2}$ \cite{Xu13,Dhaka13} and SrCo$_{2}$As$_{2}$ \cite{Pandey13} reveal a complex multi-band Fermi surface and, unlike the iron arsenides, no clear nesting features exist that might suggest an instability towards AFM ordering.

In this Letter, we report the remarkable discovery that SrCo$_{2}$As$_{2}$ is near an instability to stripe AFM order, not ferromagnetism, i.e.\ it adopts the same magnetic state as found in the tetragonal phase of iron arsenide-based parent and superconducting compounds.  Inelastic neutron scattering (INS) was used to measure steeply dispersing and quasi-two-dimensional (2-D) paramagnetic excitations near $\textbf{Q}_{\mathrm{AFM}} = (1/2,1/2,1)$ that share many similarities to $A$Fe$_{2}$As$_{2}$, despite their dissimilar Fermi surface topologies.  One notable difference is the opposite anisotropy of the longitudinal and transverse spin correlation lengths, indicating that the nearest-neighbor magnetic exchange is FM in SrCo$_{2}$As$_{2}$, rather than AFM in $A$Fe$_2$As$_2$.  Spin-polarized band-structure calculations find a tendency for both FM and stripe AFM order, which also emphasizes the important role that competing FM interactions play in the cobalt arsenides.  This result raises important questions about the origins of stripe spin-density waves and the potential for unconventional superconductivity within the cobalt arsenides.

\begin{figure*}
\includegraphics[width=0.85\linewidth]{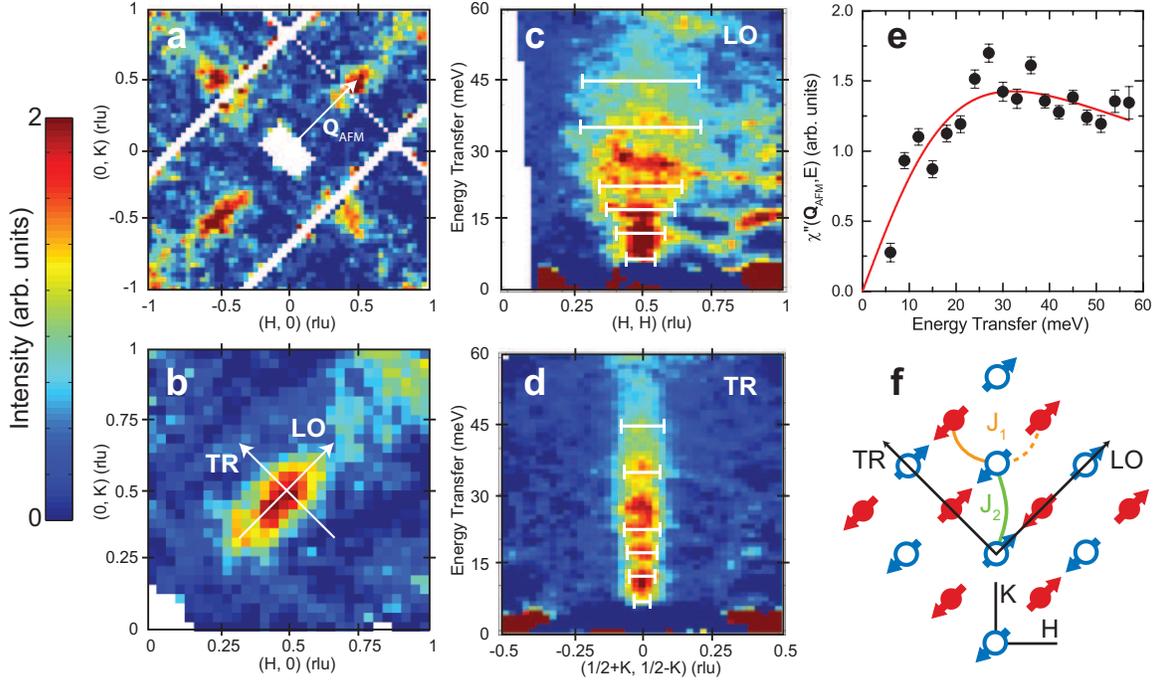}
\caption{\footnotesize   Panels (a) -- (e) show INS data on SrCo$_{2}$As$_{2}$ measured on ARCS with $E_{i}=$ 75 meV and $T=5$~K. Panel (a) shows data in the $H-K$ plane summed over 15 -- 25 meV highlighting anisotropic spin fluctuations centered at $\textbf{Q}_{\mathrm{AFM}}=(1/2,1/2)$ with $<L>$ = 1.6.  Panel (b) plots the same data as in panel (a), but symmetry-equivalent quadrants have been averaged together. Panels (c) and (d) show the energy dependence of the scattering in the LO direction [averaged over a range of $\pm$ 0.1 rlu in the TR direction] and TR direction [averaged over a range of 0.4 -- 0.6 rlu in the LO direction], respectively.  In panels (c) and (d), the white horizontal bar indicates the FWHM of single Gaussian fits to the constant energy scans, similar to those shown in Fig.~2.  (e) The imaginary part of the magnetic susceptibility centered at $\textbf{Q}_{\mathrm{AFM}}$ and averaged over the range from  $\pm$ 0.1 rlu in the TR direction and 0.35 -- 0.65 rlu in the LO direction.  The red line is a fit to a relaxational line shape, as described in the text. (f) Schematic picture of the stripe AFM order described by $\textbf{Q}_{\mathrm{AFM}}$ in the $H-K$ plane. TR (FM) and LO (AFM) directions are indicated by arrows and nearest- ($J_{1}$) and next-nearest-neighbor ($J_{2}$) exchange interactions are shown.  Full red and empty blue symbols represent Co spins on separate AFM sublattices.}
\label{fig1}
\end{figure*}
The INS measurements were carried out on the ARCS and HB3 spectrometers at the Spallation Neutron Source and High Flux Isotope Reactor, respectively, at Oak Ridge National Laboratory.  The  measurements were performed on three single crystals of SrCo$_{2}$As$_{2}$ with a total mass of approximately 2.6 g that were co-aligned to within less than 3 degrees full-width-at-half-maximum (FWHM).  Bulk electronic transport, magnetization, heat capacity, $^{75}$As NMR, and neutron diffraction measurements show no evidence for structural or magnetic phase transitions (to G-type or A-type AFM order) from room temperature down to 1.3 K, with no evidence for short-range or long-range chemical inhomogeneity \cite{Pandey13}. In the present neutron diffraction study, no detectable stripe AFM order was found in our samples down to $T = 5$ K.  The samples were mounted in the $(H,H,L)$ scattering plane and we define $\mathbf{Q} = (H,K,L) = \frac{2\pi}{a}H\hat{i} + \frac{2\pi}{a}K\hat{j} + \frac{2\pi}{c}L\hat{k}$ in reciprocal lattice units (rlu) as referenced to the tetragonal $I4/mmm$ unit cell with lattice constants $a=$ 3.94 \AA\ and $c=11.8$ \AA.  The ARCS measurements were performed with an incident neutron energy of 75 meV and the incident beam oriented along the crystallographic $L$-direction.  HB3 data were collected with a fixed final energy of 14.7 meV, horizontal collimation of 48$^{\prime}$-60$^{\prime}$-80$^{\prime}$-120$^{\prime}$, and graphite filters after the sample. 

Figures 1(a) -- (e) show INS data measured on ARCS that highlight the spin fluctuations in SrCo$_{2}$As$_{2}$. An isotropic and nonmagnetic background intensity has been estimated and subtracted using a procedure similar to that in Ref. \cite{Tucker12}. Figure 1(a), with data summed over an energy transfer range of $E=$ 15 -- 25 meV, shows magnetic scattering intensity centered at an in-plane wavevector of $\textbf{Q}_{\mathrm{AFM}}=(1/2,1/2)$ and symmetry related positions in the $H-K$ plane.  Thus, the intensity arises from stripe AFM spin fluctuations similar to the $A$Fe$_{2}$As$_{2}$ compounds.  The $L$-component of the scattering wavevector varies with $E$ in a time-of-flight experiment with fixed crystal geometry and has an average value of $<L>$ = 1.6 in this energy range.  Figure 1(b) displays the same data averaged over all four symmetry-related quadrants.  The spin fluctuations are clearly anisotropic in the $H-K$ plane, with a broader distribution of magnetic intensity in the longitudinal (LO) AFM direction [$(H,H)$-direction] than the transverse (TR) FM direction [$(K,-K)$-direction] through $\textbf{Q}_{\mathrm{AFM}}$ [see Fig. 1(f)].  No detectable experimental signature of ferromagnetic fluctuations was observed in the first Brillouin zone [close to the origin of reciprocal space in Fig.~1(a)] in the ARCS data.

Figures 1(c)  -- (e) show the energy dependence of the spin fluctuations.  The steepness of the spectrum is especially clear in the TR direction, shown in Fig.~1(d), and is similar to the iron-arsenides where the magnetic bandwidth reaches $\sim$250 meV \cite{Zhao09,Ewings11,Harriger11}.  On the other hand, the spectrum in the LO direction has a much larger and energy-dependent broadening, suggesting dispersive-like features. 

The energy dependence of the magnetic spectrum in the vicinity of $\textbf{Q}_{\mathrm{AFM}}$ is plotted in Fig.~1(e) as the imaginary part of the dynamical magnetic susceptibility
\begin{equation}
\chi^{\prime\prime}(\textbf{Q},E)=\frac{2\pi}{(\gamma r_{0})^2}\frac{I-I_{\mathrm{Bkg}}}{f^{2}(\textbf{Q})}[1-\textrm{exp}(-E/kT)].  
\label{eqn2}
\end{equation}
where $I-I_{\mathrm{Bkg}}$ is the background subtracted raw data, $(\gamma r_{0})^{2} =$ 290.6 mbarns Sr$^{-1}$, and $f(\textbf{Q})$ is the magnetic form factor of a Co$^{2+}$ ion.   The imaginary susceptibility is shown to be consistent with a relaxational spectrum, $\chi^{\prime\prime}(\textbf{Q}_{\mathrm{AFM}},E) \sim E\Gamma/(E^{2}+\Gamma^{2})$,  typical for nearly antiferromagnetic metals and yields a characteristic damping energy of $\Gamma=$ 32(8) meV.   Weak residual phonon signals still remain after the subtraction of estimated nonmagnetic background ($I_{\mathrm{Bkg}}$), especially at the higher $Q$ values in Fig.~1(c).  These residual phonon signals lead to sharp features in the energy dependence seen in Figs. 1(c)--(e).

The anisotropy of the spin fluctuations in the Co layer is clearly demonstrated by LO and TR cuts through the ARCS data, as shown in Fig.~2.  Several different energy cuts were fit to single Gaussian lineshapes with a FWHM of $\kappa_{\mathrm{LO}}$ and $\kappa_{\mathrm{TR}}$ rlu for the LO and TR cuts, respectively. The FWHM are represented as horizontal white bars in Figs. 1(c) and (d) . At low energies, the inverse FWHM of the constant energy cuts is related to the spin-spin correlation length [$\xi_{i} \approx a/(2\pi\kappa_{i}$)] and we find that  $\kappa_{\mathrm{LO}}=0.21(2)$ rlu and $\kappa_{\mathrm{TR}}=0.11(1)$ rlu.

\begin{figure}
\includegraphics[width=1.0\linewidth]{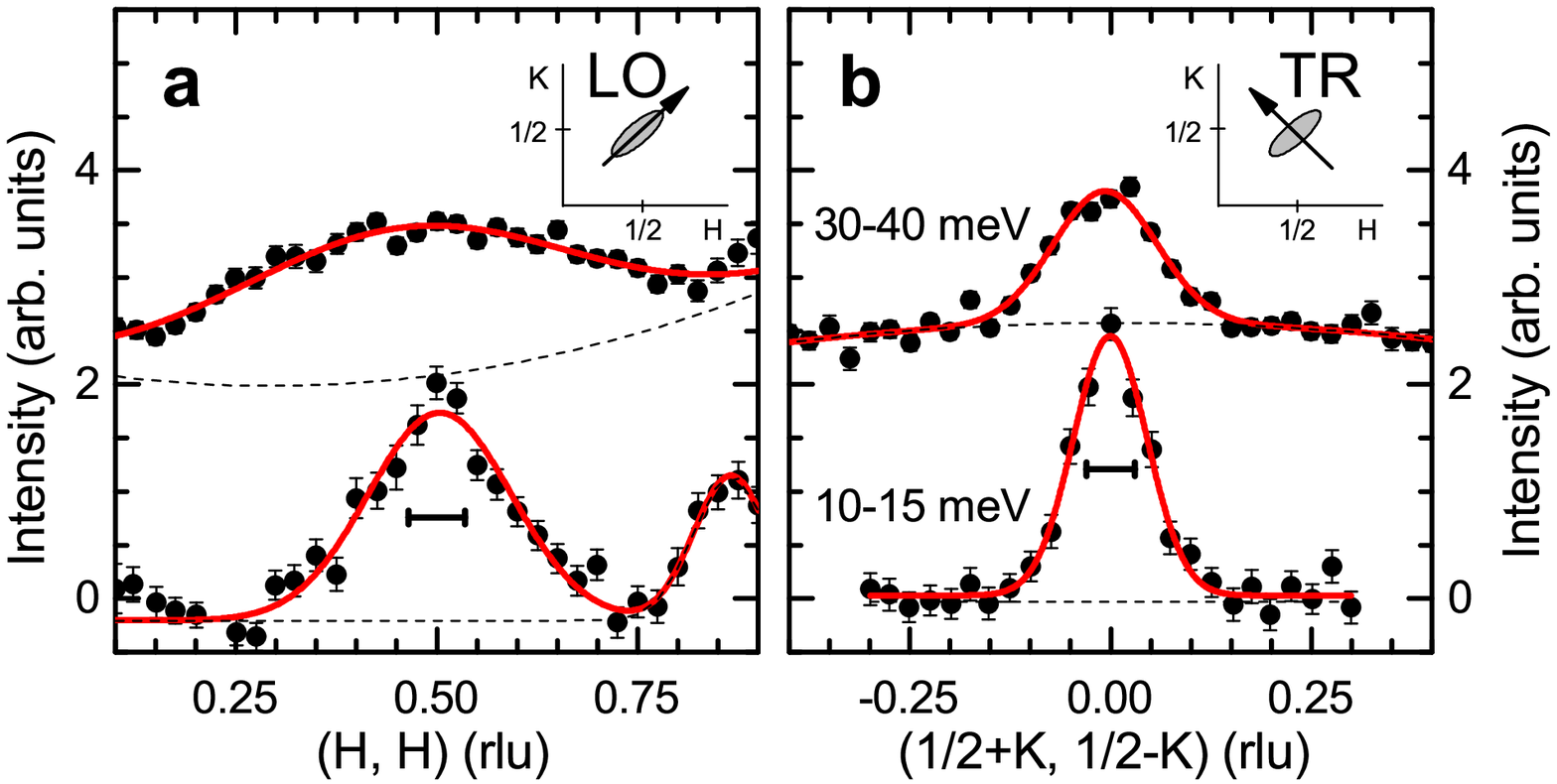}
\caption{\footnotesize (a) Longitudinal and (b) transverse cuts through $\textbf{Q}_{\mathrm{AFM}}=(1/2,1/2,<L>)$ as measured on the ARCS spectrometer and averaged over an energy range from 10 to 15 meV ($<L>$ = 1.2) and 30 to 40 meV ($<L>$ = 3.3).  The longitudinal cuts in (a) were averaged over a range of $\pm$ 0.1 rlu in the TR-direction.  The transverse cuts in (b) were averaged over a range of 0.4 -- 0.6 rlu in the LO-direction. Red lines are fits of the magnetic signal a to single Gaussian line shape and dashed lines are the fitted background.  The additional peak at 10--15 meV on the right-hand-side of (a) is a feature of the phonon background.  Horizontal bars show the estimated resolution FWHM.}
\label{fig2}
\end{figure}

Despite the likelihood that the spin fluctuations in SrCo$_2$As$_2$ are itinerant in nature, the low energy spin dynamics can be interpreted and parameterized using a Heisenberg model with nearest- ($J_{1}$) and next-nearest-neighbor ($J_{2}$) exchange interactions, as shown in Fig.~1(f).  A similar approach has been used extensively to describe the spin dynamics in the iron arsenides \cite{Zhao09, Ewings11,Harriger11}.  Within the Heisenberg model, the anisotropy of correlation lengths in the paramagnetic phase ($\eta$) is related to the ratio of $J_{1}$ and $J_{2}$ \cite{Diallo10},
\begin{equation}
\eta=\frac{\xi_{\mathrm{LO}}^{2}-\xi_{\mathrm{TR}}^{2}}{\xi_{\mathrm{LO}}^{2}+\xi_{\mathrm{TR}}^{2}} = \frac{\kappa_{\mathrm{TR}}^{2}-\kappa_{\mathrm{LO}}^{2}}{\kappa_{\mathrm{TR}}^{2}+\kappa_{\mathrm{LO}}^{2}} =\frac{J_{1}}{2J_{2}}.
\label{eqn1}
\end{equation}
From this relation, we obtain $\eta=-0.56(18)$ which implies that $|J_{1}| \approx |J_{2}|$.  Since $J_{2} > 0$ (AFM exchange) for stripe AFM correlations, the negative value of $\eta$ indicates that $J_{1} < 0$ (FM exchange).  A ferromagnetic $J_{1}$ results in a shorter correlation length along the LO direction, as it destabilizes the AFM nearest-neighbor correlations [see Fig.~1(f)]. This anisotropy can be compared to that of the parent and electron-doped $A$Fe$_{2}$As$_{2}$ materials where $\eta \approx 0.5$ \cite{Tucker12}, i.e.\ the anisotropy is opposite and $J_{1}$ is AFM for the iron-based compounds.  It is interesting to note that the spin fluctuation anisotropy is predicted to become negative ($\eta < 0$) in hole-doped $A$Fe$_{2}$As$_{2}$ \cite{Park10}, which has subsequently been observed in Ba$_{1-x}$K$_{x}$Fe$_{2}$As$_{2}$ \cite{Zhang11}. Here, we show that $\eta < 0$ also in SrCo$_{2}$As$_{2}$ which, based on ARPES measurements, may be described as a heavily electron-doped iron-compound \cite{Xu13,Dhaka13,Pandey13}.  

HB3 data were used to establish the $L$-dependence and the $T$-dependence of the spin fluctuations.  Figures 3(a) and (b) show $\chi^{\prime\prime}$ at $E=$ 7.5 meV along the LO $(H,H,1)$-direction and the $L$-direction at $(1/2,1/2,L)$. The data are reported in absolute units of $\mu_{\mathrm{B}}^{2}$ eV$^{-1}$ f.u.$^{-1}$ after subtraction of a fitted background function and calibration to the integrated intensity of transverse acoustic phonons in the (2, 2, 0) zone. We find that the absolute level of magnetic intensity at low energies is $\approx$ 3 times smaller than that found in the normal state of paramagnetic BaFe$_{1.85}$Co$_{0.15}$As$_{2}$ \cite{Inosov09}.  We note that the relative variation of sample absorption due to Co-absorption was corrected for in Fig. 3(b) \cite{absorption}.  However, an overall absorption correction was not attempted, so phonon calibrations of the absolute intensity are subject to potential errors of ~30\%.  Figure 3(b) indicates that $L$-dependent modulations are weak at $T=5$~K but are peaked at $L=$ \textit{odd}, which suggests a weak AFM interaction between the Co layers.  Thus, the AFM wavevector $\textbf{Q}_{\mathrm{AFM}}=(1/2,1/2,1)$ is identical to that of $A$Fe$_{2}$As$_{2}$.   Longitudinal and $L$ scans performed at $T=5$, 100, and 200~K in Figs. 3(a) and (b) show that the spin correlations are severely suppressed at 100~K and not detectable at 200~K.

\begin{figure}
\includegraphics[width=1.0\linewidth]{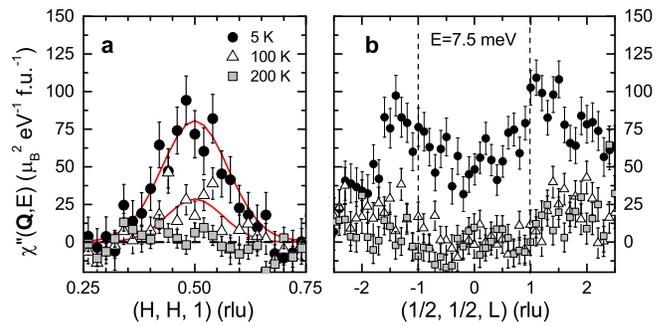}
\caption{\footnotesize Imaginary part of the magnetic susceptibility $\chi^{\prime\prime}(\textbf{Q},E)$ obtained from HB3 data at 5, 100, and 200 K and 7.5 meV (a) in the longitudinal direction along $(H,H,1)$ and (b) along the $(1/2,1/2,L)$-direction perpendicular to the Co layers.  Red lines  in (a) are Gaussian fits to the data.  There is no detectable magnetic scattering at 200 K.}
\label{fig3}
\end{figure}

We now turn to a discussion of the possible origin of stripe spin fluctuations in SrCo$_{2}$As$_{2}$.  The stripe AFM ordering in the iron arsenides is driven by Fermi surface nesting of electron and hole pockets.  However, ARPES measurements on BaCo$_{2}$As$_{2}$  \cite{Xu13, Dhaka13} and SrCo$_{2}$As$_{2}$ \cite{Pandey13} show a more complex Fermi surface and it is not obvious if a strong nesting condition exists that would support an itinerant spin-density wave description.  To better understand the possibility for Fermi surface-driven magnetism in SrCo$_{2}$As$_{2}$, we performed DFT calculations in both LDA \cite{Perdew92} and generalized gradient (GGA) approximations \cite{Perdew96} employing a full-potential linear augmented plane wave (FPLAPW) code \cite{Blaha01}. To obtain a self-consistent charge density, we used $R_{\mathrm{MT}}K_{\mathrm{max}} = 9.0$ with muffin-tin radii ($R_{\mathrm{MT}}$) of 2.3, 2.1, 2.1 a.u. for Sr, Co and As, respectively. 828 \textbf{k}-points were selected in the irreducible Brillouin zone and the calculations were iterated to reach the total energy convergence criterion of 0.01 mRy/primitive cell. Starting from experimental lattice parameters [$a=3.9471(4)$ \AA\ and $c=11.801(1)$ \AA\, and $z_{\mathrm{As}}=0.3588$] \cite{Pandey13}, we optimized the $c/a$ ratio and unit-cell volume to obtain the parameters that gave minimum total energy. Arsenic-atom positions were relaxed until the forces on As atoms were smaller than 0.1 mRy/a.u., which gave $z_{\mathrm{As}}=0.35146$ (LDA), and 0.35618 (GGA). For the $\chi(\textbf{q})$ calculations, the whole reciprocal unit cell was divided into $160\times160\times160$ parallelepipeds that resulted in 34061 \textbf{k}-points.

The LDA calculations are used to reveal wavevectors where the generalized static susceptibility $\chi(\textbf{q})$ is a maximum, signaling a tendency towards magnetic ordering at that \textbf{q}.  Similar calculations in the iron arsenides have a maximum in the susceptibility at the nesting vector for stripe AFM order \cite{Mazin08, Goldman08, Yaresko09}.  Figure 4 shows LDA calculations of $\chi(\textbf{q})$ for SrCo$_{2}$As$_{2}$ which indicate that both FM or A-type order [$\textbf{q}=(0,0,L)$] and stripe AFM order [$\textbf{q}=(1/2,1/2,L)$] with $L=0$ or 1 are preferable.  The use of experimental versus relaxed lattice parameters and As $z$-position strongly affects $\chi(\textbf{q})$, suggesting that strong magnetoelastic interactions may be present and orbital matrix elements must be included in order to establish the true magnetic ground state from DFT.  The near degeneracy of different magnetic states in SrCo$_{2}$As$_{2}$ is also demonstrated by GGA calculations of the total energy for non-magnetic, ferromagnetic, A-type AFM (FM Co layers with AFM coupling between layers), and stripe AFM ground states, as shown in the inset of Fig.~4(d).  For large $c/a$ ratios, the system tends towards a non-magnetic ground state whereas, for the experimentally observed $c/a$ ratio, ferromagnetic/A-type order is slightly preferred over stripe AFM order.  

\begin{figure}
\includegraphics[width=1.0\linewidth]{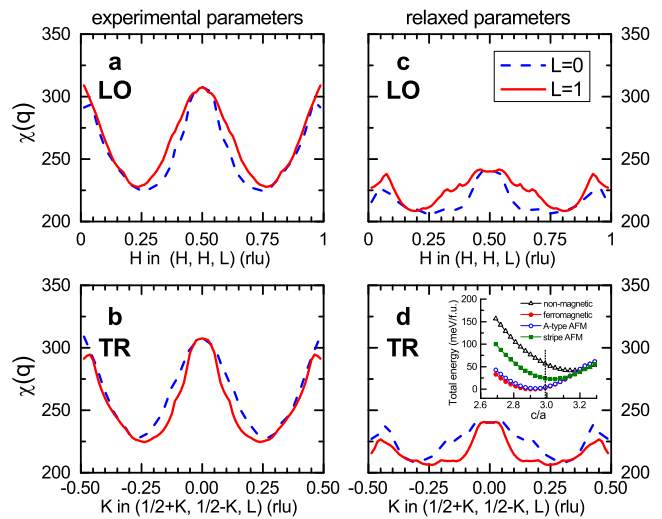}
\caption{\footnotesize DFT-LDA calculations of the \textbf{q}-dependent generalized static susceptibility, $\chi(\textbf{q})$, for SrCo$_{2}$As$_{2}$ along the (a),(c) longitudinal  and (b),(d) transverse directions at both $L=0$ (blue lines) and $L=1$ (red lines).  The calculations in (a) and (b) were performed with the experimental crystallographic parameters and in (c) and (d) were performed with lattice parameters relaxed to minimize the electronic total energy. The inset of (d) shows the GGA calculations of the total energy for different magnetic ground states.  The vertical dashed line is the experimental $c/a$ ratio.}
\label{fig4}
\end{figure}

To better understand a possible connection between stripe AFM in observed iron and cobalt arsenides, we now discuss studies of the evolution of the spin fluctuations, electronic band structure, and superconductivity on the Fe-rich end of the phase diagram of Ba(Fe$_{1-x}$Co$_{x}$)$_{2}$As$_{2}$ into the overdoped region.  Neutron scattering data indicate a complete absence of low-energy stripe spin fluctuations for $x=0.14$ \cite{Sato09,Sato11}.  The disappearance of stripe spin fluctuations can be reconciled with ARPES data showing a Lifshitz transition where the electron and hole pockets that characterize BaFe$_{2}$As$_{2}$ evolve into two mismatched electron pockets at about $x=0.20$ \cite{Liu11}.  This change in the band structure and the absence of spin fluctuations in the overdoped region also coincides with the disappearance of superconductivity.  Evidently, the stripe spin fluctuations reappear in SrCo$_{2}$As$_{2}$ with an energy scale similar to the iron arsenides. Taken together, the stripe AFM spin correlations are completely suppressed, and then restabilized, with increasing Co composition.  The reason for this behavior is likely the continuous transformation of the Fermi surface due to change in the chemical potential, although it is also possible that chemical disorder is a contributor to the suppression of stripe magnetism at intermediate compositions.  A recent investigation reporting the reestablishment of AF order with heavy electron doping of LaFeAsO$_{1-x}$H$_{x}$ shares many similarities to our current study \cite{Fujiwara13}.

Ultimately, it is worth considering whether cobalt arsenide based compounds that are tuned by chemical substitution or applied pressure will harbor unconventional superconductivity.  Certainly, the similarity of the spin fluctuations to the iron arsenides makes such an expectation possible.  However, the reciprocal space anisotropy of the spin fluctuations indicates that strong FM nearest-neighbor interactions are present in the cobalt arsenides. The prospect of incipient FM in the $A$Co$_{2}$As$_{2}$ system is supported by magnetization \cite{Cheng12,Ying12,Anand13} and neutron scattering studies \cite{Goldman13} in the collapsed tetragonal phase of CaCo$_{2}$As$_{2}$ (with $c/a=2.6$) that describe A-type AFM order below $T_{\mathrm{N}}$ = 53 to 76 K (depending on sample preparation). Our DFT calculations of the generalized static susceptibility also highlight the relative importance of competing FM interactions in the cobalt arsenides.  While unconventional superconductivity may be lurking in the cobalt arsenides, the prevalence of FM nearest-neighbor interactions and, potentially, competing FM spin fluctuations, may introduce pair-breaking that suppresses $s$-wave superconductivity and/or lead to a $p$-wave triplet superconducting state.  

The authors would like to thank P. C. Canfield, R. M. Fernandes, A. Kaminski, and A. S. Sefat for useful discussions.  The work at Ames Laboratory was supported by the U.S. Department of Energy, Office of Basic Energy Sciences, Division of Materials Sciences and Engineering under Contract No. DE-AC02-07CH11358. Work at Oak Ridge National Laboratory is supported by the U.S. Department of Energy, Office of Basic Energy Sciences, Scientific User Facilities Division.


\begin{thebibliography}{30}

\bibitem{Johnston10} D. C. Johnston,  Adv. Phys. \textbf{59}, 803 (2010).

\bibitem{Sato09} K. Matan, R. Morinaga, K. Iida, and T. J. Sato,  Phys. Rev. B \textbf{79}, 054526 (2009).

\bibitem{Sato11} T. J. Sato, K. Matan, S. Ibuka, R. Morinaga, Songxue Chi, J. W. Lynn, A. D. Christianson, and M. D. Lumsden,  Phys. Rev. B \textbf{83}, 059901 (2011).

\bibitem{Sefat08} Athena S. Sefat, Rongying Jin, Michael A. McGuire, Brian C. Sales, David J. Singh, and David Mandrus, Phys. Rev. Lett. \textbf{101}, 117004 (2008).

\bibitem{Leithe08} A. Leithe-Jasper, W. Schnelle, C. Geibel, and H. Rosner, Phys. Rev. Lett. \textbf{101}, 207004 (2008).

\bibitem{Ni08} N. Ni, M. E. Tillman, J.-Q. Yan, A. Kracher, S. T. Hannahs, S. L. Bud'ko, and P. C. Canfield,  Phys. Rev. B \textbf{78}, 214515 (2008).

\bibitem{Sefat09} A. S. Sefat, D. J. Singh, R. Jin, M. A. McGuire, B. C. Sales, and D. Mandrus,  Phys. Rev. B \textbf{79}, 024512 (2009).

\bibitem{Pandey13} Abhishek Pandey, D. G. Quirinale, W. Jayasekara, A. Sapkota, M. G. Kim, R. S. Dhaka, Y. Lee, T. W. Heitmann, P. W. Stephens, V. Ogloblichev, A. Kreyssig, R. J. McQueeney, A. I. Goldman, Adam Kaminski, B. N. Harmon, Y. Furukawa, and D. C. Johnston,  Phys. Rev. B \textbf{88}, 014526 (2013).

\bibitem{Xu13} N. Xu, P. Richard, A. van Roekeghem, P. Zhang, H. Miao, W. L. Zhang, T. Qian, M. Ferrero, A. S. Sefat, S. Biermann, and H. Ding,  Phys. Rev. X \textbf{3}, 011006 (2013).

\bibitem{Dhaka13} R. S. Dhaka, Y. Lee, V. K. Anand, D. C. Johnston, B. N. Harmon, and Adam Kaminski, Phys. Rev. B \textbf{87}, 214516 (2013).

\bibitem{Tucker12} G. S. Tucker, R. M. Fernandes, H. F. Li, V. Thampy, N. Ni, D. L. Abernathy, S. L. Bud'ko, P. C. Canfield, D. Vaknin, J. Schmalian, and R. J. McQueeney,  Phys. Rev. B \textbf{86}, 024505 (2012).

\bibitem{Zhao09} Jun Zhao, D. T. Adroja, Dao-Xin Yao, R. Bewley, Shiliang Li, X. F. Wang, G. Wu, X. H. Chen, Jiangping Hu, and Pengcheng Dai,  Nat. Phys. \textbf{5}, 555 (2009).

\bibitem{Ewings11} R. A. Ewings, T. G. Perring, J. Gillett, S. D. Das, S. E. Sebastian, A. E. Taylor, T. Guidi, and A. T. Boothroyd,  Phys. Rev. B \textbf{83}, 214519 (2011).

\bibitem{Harriger11} L. W. Harriger, H. Q. Luo, M. S. Liu, C. Frost, J. P. Hu, M. R. Norman, and Pengcheng Dai,  Phys. Rev. B \textbf{84}, 054544 (2011).

\bibitem{Diallo10} S. O. Diallo, D. K. Pratt, R. M. Fernandes, W. Tian, J. L. Zarestky, M. Lumsden, T. G. Perring, C. L. Broholm, N. Ni, S. L. Bud'ko, P. C. Canfield, H. F. Li, D. Vaknin, A. Kreyssig, A. I. Goldman, and R. J. McQueeney,  Phys. Rev. B \textbf{81}, 214407 (2010).

\bibitem{Park10} J. T. Park, D. S. Inosov, A. Yaresko, S. Graser, D. L. Sun, Ph Bourges, Y. Sidis, Yuan Li, J. H. Kim, D. Haug, A. Ivanov, K. Hradil, A. Schneidewind, P. Link, E. Faulhaber, I. Glavatskyy, C. T. Lin, B. Keimer, and V. Hinkov,  Phys. Rev. B \textbf{82}, 134503 (2010).

\bibitem{Zhang11} Chenglin Zhang, Meng Wang, Huiqian Luo, Miaoyin Wang, Mengshu Liu, Jun Zhao, D. L. Abernathy, T. A. Maier, Karol Marty, M. D. Lumsden, Songxue Chi, Sung Chang, Jose A. Rodriguez-Rivera, J. W. Lynn, Tao Xiang, Jiangping Hu, and Pengcheng Dai,  Sci. Rep. \textbf{1}, 115 (2011).

\bibitem{Inosov09} D. S. Inosov, J. T. Park, P. Bourges, D. L. Sun, Y. Sidis, A. Schneidewind, K. Hradil, D. Haug, C. T. Lin, B. Keimer, and V. Hinkov, Nat. Phys. \textbf{6}, 178 (2009).

\bibitem{absorption} The sample absorption was estimated by measurement of the attenuation of the incoherent elastic scattering signal at the identical sample positions corresponding to each inelastic scan. This absorption function was then divided into the inelastic data scans to make the correction.

\bibitem{Perdew92} J. P. Perdew and Y. Wang,  Phys. Rev. B \textbf{445}, 13244 (1992).

\bibitem{Perdew96} J. P. Perdew S. Burke and M. Ernzerhof  Phys. Rev. Lett. \textbf{77}, 3865 (1996).

\bibitem{Blaha01} P. Blaha, K. Schwarz, G.K.H. Madsen, D. Kvasnick, and J. Luitz, WIEN2k, \textit{An Augmented Plane Wave + Local Orbitals Program for Calculation Crystal Properties} (K. Schwarz, TU Wien, Austria, 2001)  ISBN 3-9501031-1-2.

\bibitem{Mazin08} I. I. Mazin, D. J. Singh, M. D. Johannes, and M. H. Du,  Phys. Rev. Lett. \textbf{101}, 057003 (2008).

\bibitem{Goldman08} A. I. Goldman, D. N. Argyriou, B. Ouladdiaf, T. Chatterji, A. Kreyssig, S. Nandi, N. Ni, S. L. Bud'ko, P. C. Canfield, and R. J. McQueeney, Phys. Rev. B \textbf{78}, 100506 (2008).

\bibitem{Yaresko09} A. N. Yaresko, G. Q. Liu, V. N. Antonov, and O. K. Andersen,  Phys. Rev. B \textbf{79}, 144421 (2009).

\bibitem{Liu11} C. Liu, A. D. Palczewski, R. S. Dhaka, T. Kondo, R. M. Fernandes, E. D. Mun, H. Hodovanets, A. N. Thaler, J. Schmalian, S. L. Bud'ko, P. C. Canfield, and A. Kaminski, Phys. Rev. B \textbf{84}, 020509 (2011).

\bibitem{Fujiwara13} N. Fujiwara, S. Tsutsumi, S. Iimura, S. Matsuishi, H. Hosono, Y. Yamakawa, and H. Kontani,  Phys. Rev. Lett. \textbf{111}, 097002 (2013).

\bibitem{Cheng12} B. Cheng, B. F. Hu, R. H. Yuan, T. Dong, A. F. Fang, Z. G. Chen, G. Xu, Y. G. Shi, P. Zheng, J. L. Luo, and N. L. Wang,  Phys. Rev. B \textbf{85}, 144426 (2012).

\bibitem{Ying12} J. J. Ying, Y. J. Yan, A. F. Wang, Z. J. Xiang, P. Cheng, G. J. Ye, and X. H. Chen,  Phys. Rev. B \textbf{85}, 214414 (2012).

\bibitem{Anand13} V. K. Anand, \textit{et al.} (unpublished).

\bibitem{Goldman13} J. H. Soh, \textit{et al.} (unpublished).

\end{thebibliography}
\end{document}